\documentclass[preprint]{revtex4-1}
\usepackage{txfonts}
\usepackage{amsfonts}
\usepackage{mathrsfs}
\usepackage{mathbbold}
\usepackage{graphicx}% Include figure files
\usepackage{subfigure}
\usepackage{dcolumn}% Align table columns on decimal point
\usepackage{bm}% bold math
\newcommand{\ds}{\displaystyle}

\begin{document}
\title[Parametrization of the transfer matrix]{Parametrization of the
transfer matrix: for one-dimensional Anderson model with diagonal
disorder}

\author{Kai Kang}
\affiliation{School of Physics, Peking University, Beijing 100871,
China} \email{colinkk@pku.edu.cn}
\author{Shaojing Qin}
\affiliation{Institute of Theoretical Physics, Chinese Academy of
Sciences, P.O. Box 2735, Beijing 100080, China}
\author{Chuilin Wang}
\affiliation{China Center of Advanced Science and Technology, P. 0.
Box 8730, Beijing 100080, China}

\begin{abstract}
In this paper, we developed a new parametrization method to
calculate the localization length in one-dimensional Anderson model
with diagonal disorder. This method can avoid the divergence
difficulty encountered in the conventional methods and
significantly save computing time as well.
\end{abstract}

\maketitle

\section{Introduction}
\label{sec:intro}

The tight-binding Anderson model \cite{PAnderson:1958} is a standard
model for disordered systems. It predicted that with random on-site
energies whose strength of disorder is above a critical value, the
electrons are localized in a certain spatial region. The transition
from metal to insulator by increasing the strength of disorder is
called Anderson transition. According to the single-parameter
scaling theory \cite{EAbrahams:1979}, there is no metallic regime in
the disordered system with dimension $d\leq2$
\cite{PMarkos:2006,FEvers:2008}. With long-range correlated
disorder, even in one-dimensional systems, there can exist extended
electronic states and a true Anderson transition with mobility edges
was observed \cite{FdeMoura:1998}.

One-dimensional systems are the simplest case to study. There were a
number of review articles on the problem of localization in
one-dimensional systems
\cite{KIshii:1973,AAbrikosov:1978,PErdos:1982,AGogolin:1982}.
Recently, articles about the general analytical methods
\cite{ARodriguez:2006}, anomalous behaviors at band center
\cite{LDeych:2003,HSchomerus:2003}, ensemble-averaged conductance
fluctuations \cite{MHilke:2008}, and discrete Anderson nonlinear
Schr\"{o}dinger equations \cite{IMata:2009} were appeared in the
major journals, indicating the topic is still a hot one.

One of the important one and quasi-one dimensional real systems is
the DNA molecules. The topics of charge transport in DNA and the
feasibility of constructing DNA-based devices, have kindled a heated
debate within the scientific community \cite{DPorath:2005}. Various
theoretical models about charge transport in DNA were generalized
from the original one-dimensional tight-binding model of Anderson
\cite{PAnderson:1958}
\begin{equation}
H=\sum_i\epsilon_ic^\dagger_ic_i+\sum_{j\neq i}t_{ji}c^\dagger_jc_i,
\end{equation}
where $\epsilon_i$ is the $i$th on-site energy given by a random
distribution, and $t_{ji}$ is the hopping energy from the $i$th site
to the $j$th site. The one-dimensional Anderson model with diagonal
disorder (all $t_{ji}$ equal to $t$) expressed in terms of the
Schr\"{o}dinger equation is
\begin{equation}
\psi_{i-1}+\psi_{i+1}=(E-\epsilon_i)\psi_i, \label{eq:SE}
\end{equation}
where $\psi_i$ is the wave function on the $i$th site, $E$ is the
eigenenergy. The hopping energy $t$ has been set as energy unit.
Generalizations to more realistic models of DNA molecules take into
account of correlations, random hopping energies, coupled
multichains, and so on
\cite{VBagci:2007,AGuo:2007,XLiu:2007,ARodriguez:2006,
GXiong:2005,MNdawana:2004,WZhang:2004,JHeinrichs:2002,AEilmes:1998}.

Of the many approaches which have been developed for numerical
simulations of disordered systems, the transfer matrix method has
proved the most productive \cite{AMacKinnon:2003,JPendry:1994}.
Introducing a two component vector $\Psi_i=(\psi_i \;\;
\psi_{i-1})^t$, Eq.~(\ref{eq:SE}) can be written as
\begin{equation}
\Psi_{i+1}=\left(\begin{array}{c} \psi_{i+1} \\ \psi_i
\end{array}\right)= \left(\begin{array}{cc} E-\epsilon_i & -1 \\ 1 & 0
\end{array}\right) \left(\begin{array}{c} \psi_i \\ \psi_{i-1}
\end{array}\right)=\mathbf{T}_i\Psi_i, \label{eq:tm}
\end{equation}
where $\mathbf{T}_i$ is the so-called transfer matrix. Here only the
diagonal disorder is considered, in which case the transfer matrix
is symplectic. With transfer matrix we can get the wavefunction
$\psi_L$ on site-$L$ propagating from site-1 with wavefunction
$\psi_1$ for arbitrary $L$ in principle by
$\Psi_L=\mathbf{T}_L\mathbf{T}_{L-1}\cdots\mathbf{T}_1\Psi_1$.

Traditional transfer matrix method has to calculate detailed
physical properties for a long chain, then to use scaling technique
to reveal the properties under the thermodynamic limit; and it
requires a stabilization procedure to overcome the overflow problem,
typically about every twelve iteration \cite{AMacKinnon:2003}. In
this paper we propose a parametrization method to deal with the
transfer matrix of one-dimensional Anderson model with uncorrelated
diagonal disorder. With this method, we directly calculate in the
localization regime under the thermodynamic limit; and it is easy to
calculate the localization length for arbitrary strength of
disorder. Particularly, it is easier to calculate moderate disorder
than both weak and strong disorder, in which cases it needs more
numerical techniques to obtain the results in our method.

\section{Parametrization of the transfer matrix}

The transfer matrix $\mathbf{T}_i$ in Eq.~(\ref{eq:tm}) is a
symplectic matrix, which satisfies the condition
$\mathbf{A}^t\mathbf{\Omega}\mathbf{A}=\mathbf{\Omega}$, where
$\mathbf{A}^t$ is the transpose of a $2n\times 2n$ matrix
$\mathbf{A}$, and $\mathbf{\Omega}$ is a fixed nonsingular,
skew-symmetric matrix. Typically $\mathbf{\Omega}$ is chosen to be a
block matrix
\begin{equation}
\mathbf{\Omega}=\left[\begin{array}{cc}\mathbbold{0} & \mathbbold{1} \\
-\mathbbold{1} & \mathbbold{0} \end{array}\right],
\end{equation}
where $\mathbbold{0}$ and $\mathbbold{1}$ are $n\times n$ zero
matrix and unit matrix respectively. It is easily shown from the
definition that the transpose of a symplectic matrix is also a
symplectic matrix.

Let $\mathbf{M}_L=\mathbf{T}_1\mathbf{T}_2\cdots\mathbf{T}_L$ and
$v_i=E-\epsilon_i$. Since the symplectic matrices form a group, the
product of  $\mathbf{M}^t_L$ and $\mathbf{M}_L$ is symplectic and
real symmetric. Thus, it can be diagonalized by an orthogonal matrix
$\mathbf{U}(\theta_L)$,
\begin{equation}
\mathbf{U}(\theta_L)\mathbf{M}^t_L\mathbf{M}_L \mathbf{U}(-\theta_L)
=\left(\begin{array}{cc} e^{\lambda_L} & \\ & e^{-\lambda_L}
\end{array} \right),
\end{equation}
where
\begin{equation}
\mathbf{U}(\theta_L)=\left(\begin{array}{cc} \cos\theta_L &
-\sin\theta_L
\\ \sin\theta_L & \cos\theta_L \end{array} \right).
\end{equation}
The $2\times2$ symplectic matrix has a property that the two
eigenvalues are reciprocal to each other. Hence,
$\mathbf{M}^t_L\mathbf{M}_L$ can be expressed in terms of
$\lambda_L$ and $\theta_L$,
\begin{equation}
\mathbf{M}^t_L\mathbf{M}_L
=\cosh\lambda_L\mathbbold{1}+\sinh\lambda_L\left(\begin{array}{cc}
\cos2\theta_L & -\sin2\theta_L \\ -\sin2\theta_L & -\cos2\theta_L
\end{array}\right),\label{eq:para}
\end{equation}
here $\mathbbold{1}$ is a $2\times2$ unit matrix and the two
parameters $\lambda_L$ and $\theta_L$ play important roles to
parameterize  the transfer matrices.

Making use of the definition of $\mathbf{M}_{L}$,
\begin{equation}
\mathbf{M}^t_{L+1}\mathbf{M}_{L+1} = \left(\begin{array}{cc} v_{L+1}
& 1 \\ -1 & 0
\end{array}\right)\mathbf{M}^t_L\mathbf{M}_L
\left(\begin{array}{cc} v_{L+1} & -1 \\
1 & 0 \end{array}\right),
\end{equation}
and substituting the expression for $\mathbf{M}^t_L\mathbf{M}_L$ in
Eq.~(\ref{eq:para}), we can easily obtain the recursion relations
for $\lambda$ and $\theta$,
\begin{eqnarray}
\cosh\lambda_{L+1}=(1+\frac{v_{L+1}^2}{2})\cosh\lambda_L
+(\frac{v_{L+1}^2}{2}\cos2\theta_L-v_{L+1}\sin2\theta_L)
\sinh\lambda_L \label{eq:rec1}\\
\sinh\lambda_{L+1}\cos2\theta_{L+1}=[(\frac{v_{L+1}^2}{2}-1)
\cos2\theta_L-v_{L+1}\sin2\theta_L]\sinh\lambda_L+\frac{v_{L+1}^2}{2}
\cosh\lambda_L \label{eq:rec2}\\
\sinh\lambda_{L+1}\sin2\theta_{L+1}=v_{L+1}\cosh\lambda_L
+(v_{L+1}\cos2\theta_L-\sin2\theta_L)\sinh\lambda_L. \label{eq:rec3}
\end{eqnarray}

In the localization regime, the localization length is finite. For
sufficiently long chains (depending on the strength of
localization), the exponent $\lambda$ of the eigenvalue will be
approaching to infinity. Hence after dividing Eq.~(\ref{eq:rec3}) by
Eq.~(\ref{eq:rec2}) and taking the limit $\lambda\rightarrow\infty$
or $\tanh\lambda_L\rightarrow 1$, we get the recursion relation of
$2\theta$ in the localization regime,
\begin{equation}
\tan2\theta_{L+1}=\frac{v_{L+1}(1+\cos2\theta_L)-\sin2\theta_L}
{v_{L+1}^2(1+\cos2\theta_L)/2-v_{L+1}\sin2\theta_L-\cos2\theta_L}.
\label{eq:2theta}
\end{equation}
Using the basic relations of trigonometric functions, it is easy to
derive a much simpler recursion relation of $\theta$,
Eq.~(\ref{eq:theta}), from Eq.~(\ref{eq:2theta}),
\begin{equation}
\tan\theta_{L+1}=\frac{1}{v_{L+1}-\tan\theta_L}. \label{eq:theta}
\end{equation}
From Eq.~(\ref{eq:theta}) it is apparent that if $v_{L+1}=0$,
$\theta_{L+1}=\theta_L\pm\pi/2$, which is also indicated in
Eq.~(\ref{eq:2theta}). Eq.~(\ref{eq:theta}) can also be written as a
continued fraction
\begin{equation}
\tan\theta_{L}=\frac{1}{\displaystyle v_L-\frac{1}{\displaystyle
v_{L-1}-\cdots-\frac{1}{\displaystyle v_2-\tan\theta_1}}}.
\end{equation}
In this form we can see that, for each specified chain, $\theta_L$
is completely determined by the sequence of random on-site energies
and the eigenenergy $E$ in the Schr\"{o}dinger equation. We should
bear in mind that all the results we have in this paper are in the
localization regime under the thermodynamic limit. Therefore,
although we label the first site as site 1, it does not mean that
site 1 starts from the very beginning, there are sufficiently many
sites before it. Furthermore, as an initial input, the effect of
$\theta_1$ will vanish for sufficiently large $L$.

Now we derive the probability density of $\theta$ from the known
random distribution of $\epsilon$ (or $v=E-\epsilon$). And we can
calculate the localization length directly through the probability
density of $\theta$. Define $t\equiv\tan\theta$. $ $From
Eq.~(\ref{eq:theta}), we have an integral equation for the
probability density function of $t$
\begin{eqnarray}
p_c(\frac{1}{t})&=&\int_{-\infty}^\infty\int_{-\infty}^\infty
p_t(t')p_v(v)\delta[\frac{1}{t}-(v-t')]\mathrm{d}t'\mathrm{d}v \nonumber \\
&=&\int_{-\infty}^\infty p_t(t')p_v(\frac{1}{t}+t')\mathrm{d}t',
\label{eq:den_eq}
\end{eqnarray}
where $p_t(t)$, $p_c(\ds\frac{1}{t})$, and $p_v(v)$ are the
probability density functions for $\tan\theta$, $\cot\theta$, and
$v$ respectively. With the relations of the probability density
functions,
\begin{equation}
p(\theta)= {1\over\cos^2\theta}p_t(t)=
{1\over\sin^2\theta}p_c(\frac{1}{t}), \label{eq:den_rel}
\end{equation}
the integral equation of the probability density function
$p(\theta)$ becomes
\begin{equation}
p(\theta)=\frac{1}{\sin^2\theta}\int_{-\pi/2}^{\pi/2}
p(\theta')p_v(\frac{1}{\tan\theta}+\tan\theta')\mathrm{d}\theta'.\label{eq:theta_distri}
\end{equation}
If we can obtain the solution of $p(\theta)$ from
Eq.~(\ref{eq:theta_distri}), the inverse localization length is
given by Eq.~(\ref{eq:rec1})
\begin{eqnarray}
\frac{1}{\xi}&=&\gamma=\frac{1}{2}
\langle\lambda_{L+1}-\lambda_L\rangle \nonumber \\
&=&\frac{1}{2}\int_{-\pi/2}^{\pi/2}\int_{-\infty}^{\infty}
p(\theta)p_v(v)\ln(1+v^2\cos^2\theta-v\sin2\theta)\mathrm{d}v
\mathrm{d}\theta, \label{eq:length}
\end{eqnarray}
where $\gamma$ is the so-called Lyapunov exponent. The second line
is obtained when $\lambda\rightarrow\infty$, hence
$\cosh\lambda_{L+1}/\cosh\lambda_L\rightarrow\exp(\lambda_{L+1}-\lambda_L)$
and $\tanh\lambda_L\rightarrow1$. For uncorrelated disorder
considered in this paper, $p(\theta)$ and $p_v(v)$ are independent
of each other, thus the above equation is correct for sufficiently
long chains in the localization regime. In \cite{ARodriguez:2006}, a
similar expression was obtained from a different starting point for
$p_v(v)$ as a discrete distribution and the parameter $\theta$ was
chosen to be real at the beginning.

It should be mentioned that from the original Schr\"{o}dinger
equation we have
\begin{equation}
z_{L+1}=\frac{1}{v_{L+1}-z_L}, \label{eq:z}
\end{equation}
where $z_L=\psi_L/\psi_{L+1}$. This is exactly the form of
Eq.~(\ref{eq:theta}). The difference is that $z$ is a complex number
while $\tan\theta$ is real. We found that for sufficiently large $L$
with the same random sequence, the real part of $z_L$ is the same as
$\tan\theta_L$ and its imaginary part goes to zero in the
localization regime.  $z_L$ and $\tan\theta_L$ are essentially the
same quantity when $L$ is sufficiently large in the localization
regime. Thus the phase difference between $\psi_L$ and $\psi_{L+1}$
is 0 or $\pi$ and the ratio of their amplitudes is
$\mathrm{Re}(\psi_L/\psi_{L+1})\rightarrow\tan\theta_L$ for
sufficiently large $L$ in the localization regime. And the
expression for the localization length of $z$ in \cite{KIshii:1973}
can be rewritten in our symbols as the following,
\begin{eqnarray}
\gamma(E)=-\int_{-\pi/2}^{\pi/2} \mathrm{d}\theta
p(\theta)\ln\left|\tan\theta\right|, \label{eq:log_tan_gamma}
\end{eqnarray}
once we solve the probability density $p(\theta)$ from
Eq.~(\ref{eq:theta_distri}), where the dependence on the eigenenergy
$E$ is implicitly included in $p(\theta)$. The $z$ defined in
\cite{KIshii:1973} is the inverse of ours, thus we have a minus sign
in our equation. Eq.~(\ref{eq:length}) and Eq.~(\ref{eq:log_tan_gamma}) are
equivalent
\begin{eqnarray*}
& &\frac{1}{2}\iint p(\theta)p_v(v)\ln(1+v^2\cos^2\theta-v\sin2\theta)
\mathrm{d}v\mathrm{d}\theta\\
&=&\frac{1}{2}\left\{\int p(\theta)\ln\cos^2\theta\mathrm{d}\theta+\iiint \mathrm{d}\theta
\mathrm{d}v\mathrm{d}\cot\theta' p(\theta)p_v(v)\ln[1+(v-\tan\theta)^2]
\delta[\cot\theta'-(v-\tan\theta)]\right\}\\
&=&-\int p(\theta)\ln|\tan\theta|\mathrm{d}\theta.
\end{eqnarray*}
However, Eq.~(\ref{eq:length}) is more appropriate for
numerical calculation because it requires less accurate $p(\theta)$
than Eq.~(\ref{eq:log_tan_gamma}).

From the above description, we can experience benefits of our
method. We do not have to use the recursive relation
Eq.~(\ref{eq:theta}) to calculate the sequence $\theta_L$, nor
$\lambda_L$. Although the recursive relation is very simple, the
time consumption of the computation of $\theta_L$ is still the same
as in the conventional transfer matrix method within the same
accuracy. Moreover, the conventional transfer matrix method has a
problem that it has to care about the overflow problem especially
for strong disorder. This is not a problem in our method. We can
calculate arbitrary strength disorder. More importantly, we can
easily calculate moderate strength disorder, which is difficult for
analytical methods.

\section{Numerical technique}
Eq.~(\ref{eq:theta_distri}) shall be solved numerically in a
discrete matrix form by using self-consistent iterative procedure.
In principle, we can get the probability density $p(\theta)$ for any
known random distribution $p_v(v)$. In this paper we only considered
two distributions: Lorentzian and Gaussian. For the Lorentzian
distribution, there are exact analytical results \cite{KIshii:1973}
and for the Gaussian distribution, there are analytical results in
the weak and strong disorder limits \cite{FIzrailev:1998}. We will
compare our numerical results with theirs.

From the r.h.s of Eq.~(\ref{eq:theta_distri}), we can see that there
is a removable singularity at $\theta=0$ in both of the Lorentzian
and Gaussian distributions in the numerical computation. Thus we
need to use interpolation to obtain $p(\theta)$ near $\theta=0$. For
the Lorentzian distribution we use six-points interpolation and for
the Gaussian distribution we use seven-points interpolation with a
condition $p(0)=[p(\pi/2)+p(-\pi/2)]/2$. This condition can be
easily derived from Eq.~(\ref{eq:den_eq}). Let
$t''=t'+\ds{\frac{1}{t}}$, then Eq.~(\ref{eq:den_eq}) becomes,
\begin{equation}
p_c(\frac{1}{t}) =\int_{-\infty}^\infty
p_t(t''-\frac{1}{t})p_v(t'')\mathrm{d}t''. \label{eq:den_eq2}
\end{equation}
For $|t''|\ll|1/t|$, $p_t(t''-1/t)\approx p_t(-1/t)$, so
Eq.~(\ref{eq:den_eq2}) can become approximately,
\begin{equation}
p_c(\frac{1}{t}) \approx p_t(-\frac{1}{t})\int_{-\infty}^\infty
p_v(t'')\mathrm{d}t'', \label{eq:den_eq3}
\end{equation}
and we have $\ds \lim_{t \rightarrow 0^\pm}p_c(\frac{1}{t}) =
\lim_{t \rightarrow 0^\pm} p_t(-\frac{1}{t})$. This relation can be
rewritten in terms of the probability density of $\theta$ as
$p(\theta\rightarrow0^\pm)=p(\mp \pi/2)$ by using
Eq.~(\ref{eq:den_rel}). We should emphasize that for the Lorentzian
distribution this relation is not valid. The range of $\theta$ need
to use interpolation is determined by
$|\theta|<\sqrt{m\pi/N\sigma}$, where $m$ is a number we choose to
ensure that the profile of $p(\theta)$ near $\theta=0$ is smooth (we
assume this is true for nonsingular distributions $p_v(v)$), $N$ is
the number of equally spaced abscissas points where the integrands
are evaluated and $\sigma$ is the parameter of the distributions,
$p_v(v)=\ds \frac{1}{\sigma\sqrt{\pi}}\exp[-(v-E)^2/\sigma^2]$ and
$p_v(v)=\ds \frac{1}{\pi}\frac{\sigma}{\sigma^2+(v-E)^2}$,
respectively. The magnitude of $\sigma$ in the distribution function
represents the strength of disorder. With $p(\theta)$ in the hand,
we can use Eq.~(\ref{eq:length}) to calculate the localization
length numerically for cases of uncorrelated disorder.

In our method, errors come from the value $N$ we choose as the
number of discrete components of $p(\theta)$ and where
the cut-off is in the infinite range of the integration with respect
to $v$. Basically, large $E$ or small $\sigma$ need large $N$, and
large $\sigma$ needs a large range of integration and a correction
to remedy the effect of finite integration range. When $N$ is large,
it spends a long time on the computation of $p(\theta)$. In fact,
this is always the main part of the computation time no matter how
large $N$ is. When the range of integration for $v$ is large, more
CPU time is consumed to compute the integration of $v$; increasing
$N$ does not improve the result when the range of integration remain
unchanged; the result is sensitive to the range of integration of
$v$ for the Gaussian distribution, while for the Lorentzian
distribution it is not. The reason should be that the Gaussian
distribution varies more dramatically than the Lorentzian
distribution.

\section{Results and discussions}

\begin{figure}
\centering \subfigure[]{
\includegraphics[width=4cm]{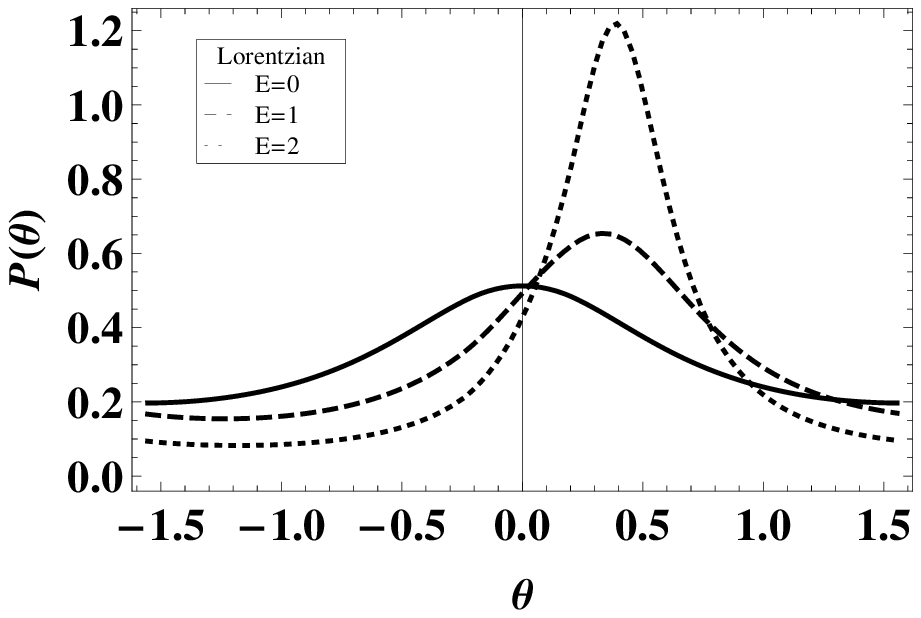}
\label{fig:figure1a}} \subfigure[]{
\includegraphics[width=4cm]{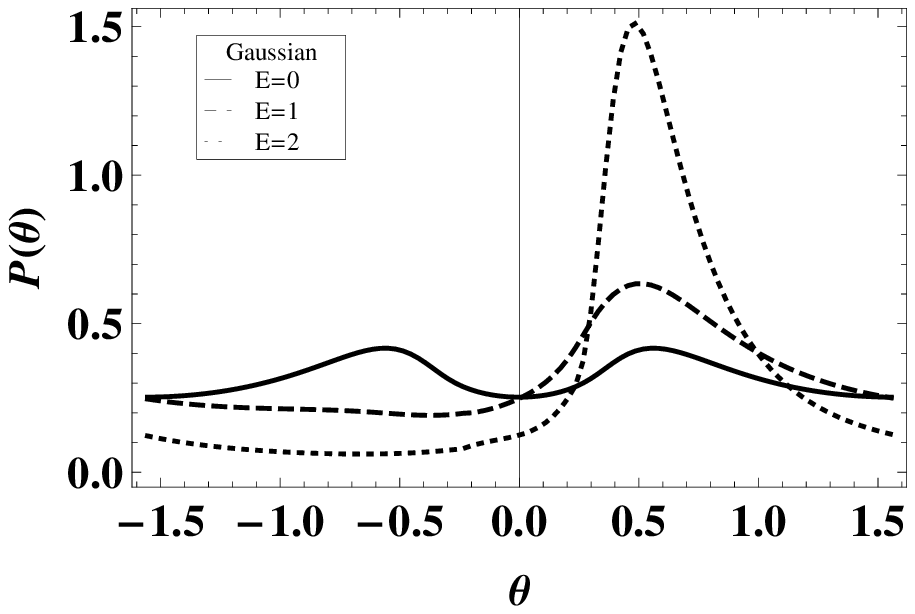}
\label{fig:figure1b}} \caption{Probability density $p(\theta)$
obtained numerically from Eq.~(\ref{eq:theta_distri}) with $E$ = 0
(solid), 1 (dashed), 2 (dotted) when $\sigma=1$ for (a) the
Lorentzian distribution $p_v(v)=
\ds\frac{1}{\pi}\frac{1}{1+(v-E)^2}$, (b) the Gaussian distribution
$p_v(v)=\ds\frac{1}{\sqrt{\pi}}\exp[-(v-E)^2]$.} \label{fig:figure1}
\end{figure}

As mentioned in the previous section, we obtained $p(\theta)$
numerically from Eq.~(\ref{eq:theta_distri}) first. We performed the
integral with respect to $v$ by using Romberg's method, and solved
the integral equation by using a discrete matrix form, where
$p(\theta)$ was written in a vector form with $N$ components. Figure
\ref{fig:figure1} shows the solutions of $p(\theta)$ for the
Lorentzian and Gaussian distributions respectively. When $E=0$,
$p(\theta)$ is symmetric respected to $\theta=0$. As $E$ becomes
large, $p(\theta)$ becomes a Dirac $\delta$-function. The position
of the peak does not change monotonically with the increase of $E$.
The salient difference of $p(\theta)$ between the two distributions
is, when $E$ is small, the Lorentzian distribution has only one
peak, while the Gaussian distribution has two.

\begin{figure}
\centering \subfigure[]{
\includegraphics[width=4cm]{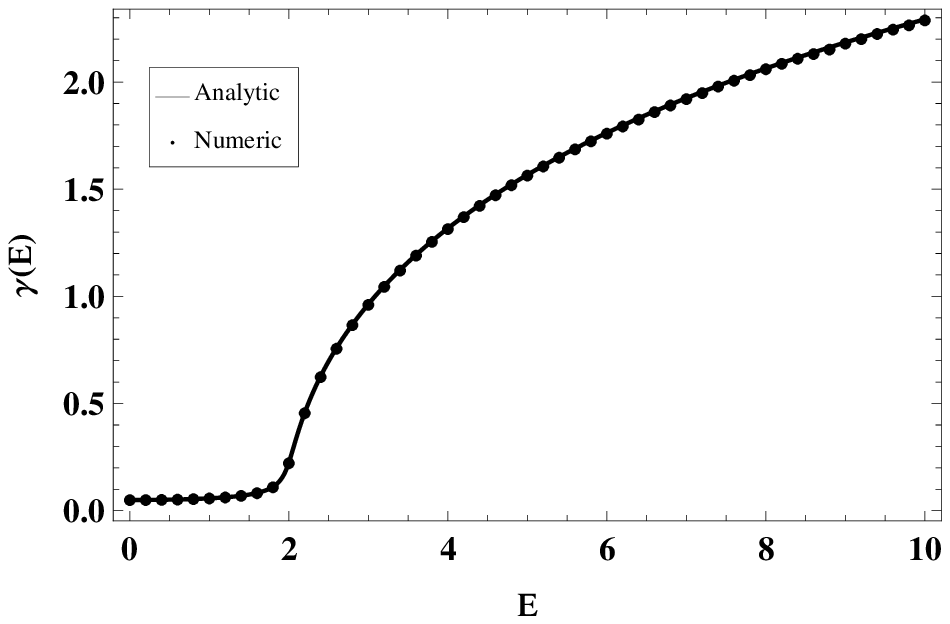}
\label{fig:figure2a}} \subfigure[]{
\includegraphics[width=4cm]{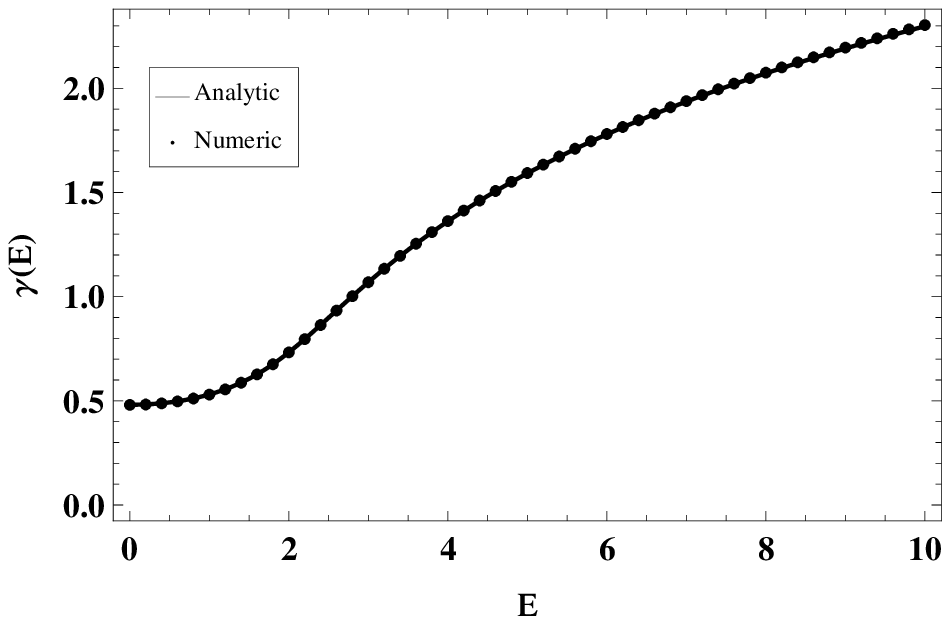}
\label{fig:figure2b}} \subfigure[]{
\includegraphics[width=4cm]{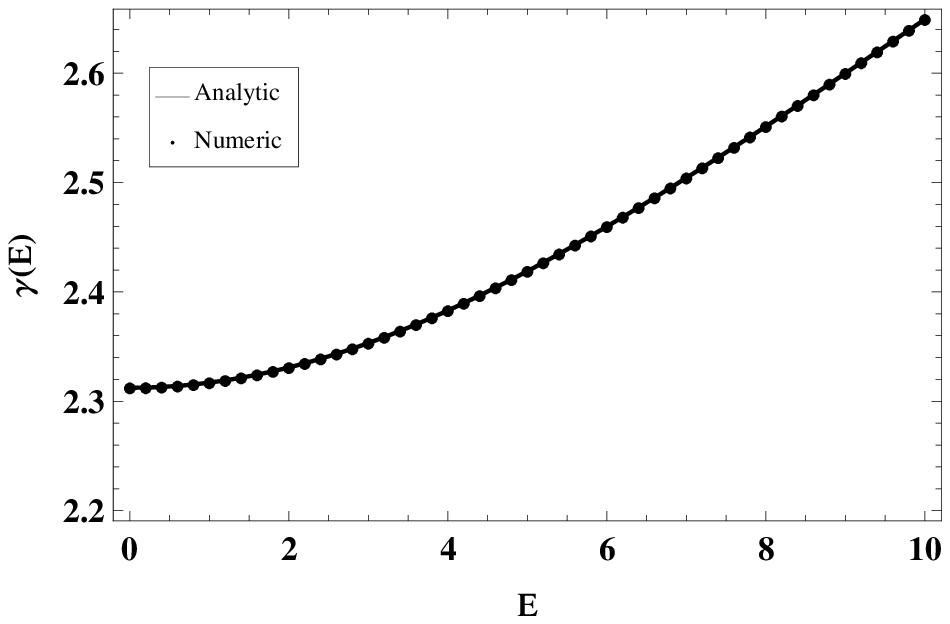}
\label{fig:figure2c}} \caption{Comparison between the analytical
and our numerical results for the Lorentzian
distribution with $\sigma$ = (a) 0.1, (b) 1, and (c) 10,
respectively.} \label{fig:figure2}
\end{figure}

After we had the probability density $p(\theta)$, the inverse
localization length $\gamma$ was calculated from
Eq.~(\ref{eq:length}) numerically. For the Lorentzian distribution,
the analytical results were already available in \cite{KIshii:1973},
\begin{equation}
\gamma(E,\sigma)=\mathrm{arccosh}\frac{\sqrt{(2+E)^2+\sigma^2}
+\sqrt{(2-E)^2+\sigma^2}}{4}. \label{eq:lor}
\end{equation}
Our numerical results for the Lorentzian distribution are shown in
Figure \ref{fig:figure2}, where we plot $\gamma$ versus $E$ with
$\sigma=0.1,~1,~10$ respectively with the comparison between our
results and the analytical results. We found that they all coincide
in excellence, and we can not see the difference from the figure. We
used the Romberg integration method to perform a numerical integral
with respect to $v$, and we made a cut-off to the infinity range of
integration, so the real range is finite from $E-10^3\sigma$ to
$E+10^3\sigma$. Because the Lorentzian distribution does not decay
rapidly to zero when $\sigma$ was large, we need to consider the
compensation of the contribution from the cut-out range. In fact,
the integral of the cut-out range can be approximately integrated
analytically, and the correction was
$4\ln|10^3\sigma\cos\theta|/10^3\pi$. This correction would
eliminate a constant difference between the numerical result and the
analytical result. For weak disorder $\sigma=0.1$ shown in Figure
\ref{fig:figure2a}, with increasing $E$, the number of mesh points
$N$ should go from 2000 to 9000 in order to obtain the right result,
otherwise the result $\gamma$ will be much smaller than the
analytical result. For strong disorder $\sigma=10$ shown in Figure
\ref{fig:figure2c}, the number of mesh points $N=2000$ with the
correction is good enough to reach the desired accuracy, no matter
how large $E$ is.

\begin{figure}
\includegraphics[width=4cm]{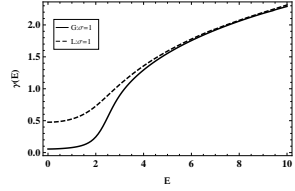}
\caption{Inverse of the localization length
$\gamma(E)$ obtained from Eq.~(\ref{eq:log_tan_gamma}) and
Eq.~(\ref{eq:length}) for the Gaussian (solid) and the Lorentzian
(dashed) distributions.} \label{fig:figure3}
\end{figure}

Numerical calculation of Eq.~(\ref{eq:log_tan_gamma}) and
Eq.~(\ref{eq:length}) gave the same result in Figure \ref{fig:figure3}.
This is one evidence of the conclusion that $z$ and $\tan\theta$ are
the same quantity when the chain is long enough in the
localization regime.
From Figure \ref{fig:figure3}, it can be seen that when in the band
$E=0$ to 2, the difference of $\gamma$ between the Lorentzian and
the Gaussian distributions is nearly a constant. It seems that they
have the same behavior in the band. When $E$ becomes large, the
$\gamma$ of the two distributions increase and go to the same value.
The reason is that when $E\rightarrow\infty$, both the Lorentzian
and Gaussian distributions become a $\delta$-function. For example,
for the Lorentzian distribution, let $y=v/E-1$,
\begin{equation}
\lim_{E\rightarrow\infty}\frac{1}{\pi}\frac{\sigma}{(v-E)^2+\sigma^2}
\mathrm{d}v=\delta(y)\mathrm{d}y,
\end{equation}
then
\begin{eqnarray}
\int_{-\infty}^\infty f(v)p_v(v)\mathrm{d}v\nonumber
&=&\int_{-\infty}^\infty f(Ey+E)\delta(y)\mathrm{d}y\nonumber\\
&=&f(E).
\end{eqnarray}
Thus if $p(\theta)$ at large $E$ are the same for different
$p_v(v)$, $\gamma(E)$ will be the same. We have checked that this is
true for the two distributions.

\begin{figure}
\centering \subfigure[]{
\includegraphics[width=4cm]{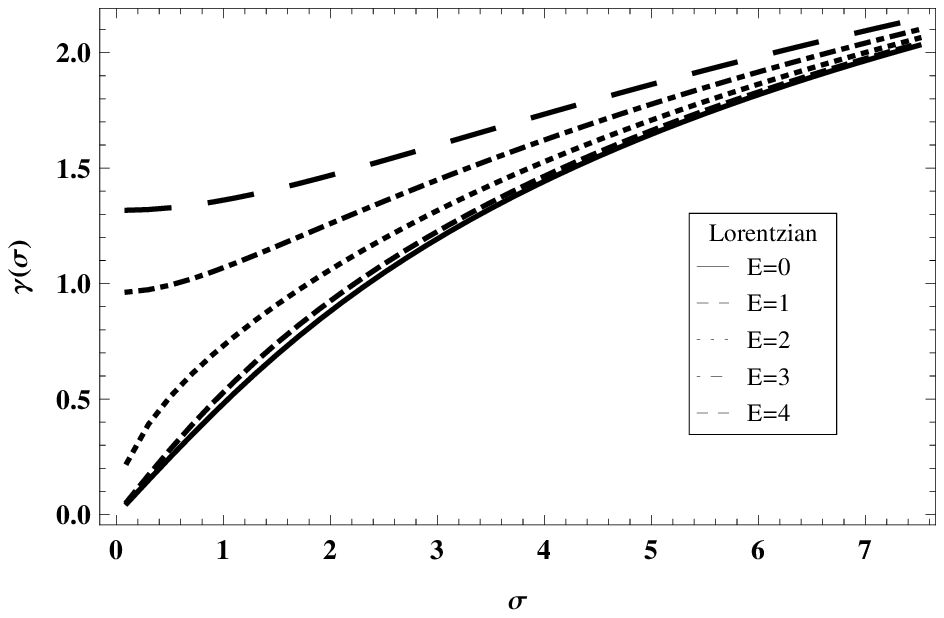}
\label{fig:figure4a}} \subfigure[]{
\includegraphics[width=4cm]{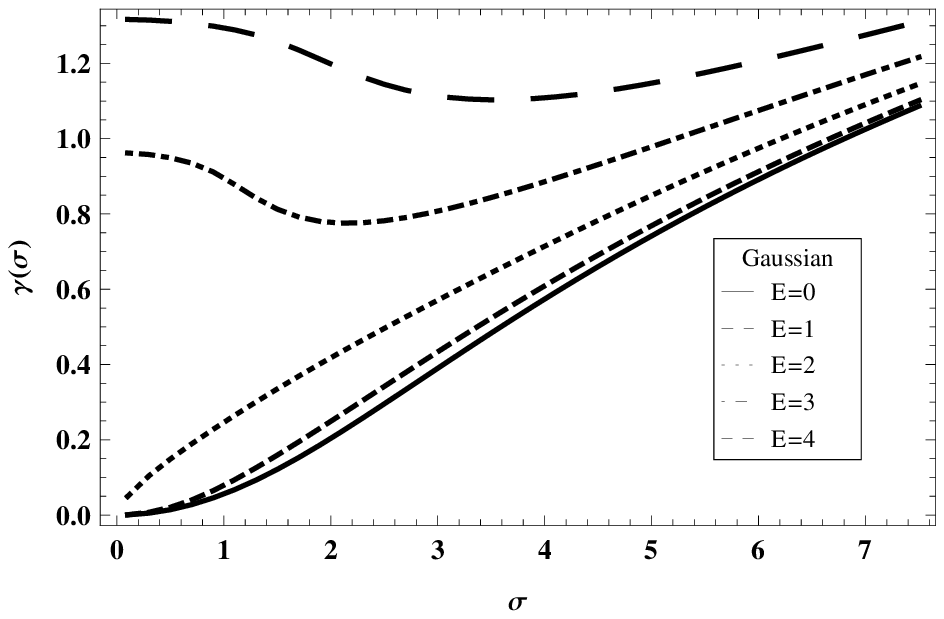}
\label{fig:figure4b}} \caption{Inverse of the localization length
$\gamma(\sigma)$ from Eq.~(\ref{eq:length}) for (a) the Lorentzian
distribution and (b) the Gaussian distribution with $E=0,1,2,3,4$
(from bottom to top).} \label{fig:figure4}
\end{figure}

In Figure \ref{fig:figure4} we plot $\gamma$ versus $\sigma$. For
the Lorentzian distribution, the results also coincide well with
Eq.~(\ref{eq:lor}). And we found that all the five lines can be
fitted well by $\gamma(\sigma)=a+b\sqrt{\sigma}+c\sigma+d\sigma^2$.
For all the five lines, $d$ is always at least one order smaller
than $b$ and $c$. For the Gaussian distribution, this is also true
when $E$ is in the band. When $E>2$ out of the band, the behavior of
the Gaussian distribution is very different and it can not be fitted
by the above formula. When $\sigma$ becomes large, $\gamma(\sigma)$
becomes the same for different $E$. This is apparent in Figure
\ref{fig:figure4a} for the Lorentzian distribution. Similar to the
derivation of large $E$, when $\sigma$ grows large, the difference
between $p_v(v)$ with different $E$ vanishes. And this can be
understood that when the width of the distribution becomes large,
the position of the peak is irrelevant.

In \cite{FIzrailev:1998} the authors provided analytical results for
the Gaussian distribution with weak and strong disorder, which are
summarized as follows,
\begin{equation}
\gamma(E,\sigma)=\left\{\begin{array}{r@{\quad,\quad}l}W^2/105.045\cdots
& E=0 \\ W^2/24(4-E^2) & 0<E<2 \\ 0.289\cdots(\delta^2)^{1/3} & E=2
\end{array} \right.. \label{eq:anal}
\end{equation}
In \cite{FIzrailev:1998}, the strength of disorder are represented
by parameters $W$ or $\delta$, which can be expressed in terms of
$\sigma$ in our paper by $W^2=6\sigma^2$ and $\delta^2=\sigma^2/2$.

For the purpose of comparison, we showed the fitting formulae of our
numerical data at weak disorder $0<\sigma<1$ in Table
\ref{table:fit_weak}.
\begin{table}
\begin{tabular}{|c||c|c|}
\hline $E$ & Lorentzian & Gaussian \\ \hline 0 & $0.49\sigma$ &
$0.056\sigma^2$ \\ \hline 1 & $0.55\sigma$ & $0.081\sigma^2$ \\
\hline 2 & $0.72\sqrt{\sigma}$ & $0.24\sigma^{2/3}$ \\ \hline 3 &
$0.96+0.11\sigma^2$ & $0.96-0.063\sigma^2$ \\ \hline 4 &
$1.32+0.045\sigma^2$ & $1.32-0.022\sigma^2$ \\ \hline
\end{tabular}
\caption{Fit for weak disorder $0<\sigma<1$}\label{table:fit_weak}
\end{table}
The ratio of the coefficients of $E=1$ and $E=0$ derived from
Eq.~(\ref{eq:anal}) is approximately 1.46 and the ratio of our
numerical results calculated from Table \ref{table:fit_weak} is
approximately 1.45. At the band edge $E=2$, our result is
$\gamma(2)\approx0.24\sigma^{2/3}$, i.e.~approximately
$0.30\delta^{2/3}$. Therefore, our results confirmed the anomalous
behavior of weak disorder at the band center $E=0$
\cite{MKappus:1981,BDerrida:1984,GCzycholl:1981} and the band edge
$E=2$ \cite{BDerrida:1984}. Furthermore, our numerical result gave
an expression for strong disorder at $E=1$,
$\gamma(\sigma)=-0.80+1.00\ln\sigma$, which is in excellent
agreement with the result of \cite{FIzrailev:1998}, where
$\gamma(\sigma)=-0.797\cdots+\ln\sigma$ when $E$ is in the band.

We noticed that for the Lorentzian distribution, the results can be
obtained from the exact expression Eq.~(\ref{eq:lor}) for small and
large $\sigma$ limits respectively. For strong disorder,
$\gamma(\sigma)=\ln\sigma$ and our fitting formulae is
$\gamma(\sigma)=1.00\ln\sigma$. It seems that the Lorentzian
distribution should not have the anomalous behavior at the band
center or band edge. However, it is apparent from Table
\ref{table:fit_weak} that at $E=2$, the Lorentzian distribution also
has an anomalous behavior similar to that of the Gaussian
distribution. On the other hand, it should be mentioned that $E=0$
may should be seen as a boundary of two bands rather than the center
of one band \cite{LDeych:2003}. As shown in Figure
\ref{fig:figure1}, the Lorentzian distribution has only one peak,
while the Gaussian distribution has two when $E$ is small.

\section{Conclusions}

In this paper, we derived a parametrization method to deal with the
transfer matrix of the one-dimensional Anderson model with diagonal
uncorrelated disorder. With this method, we directly calculated the
localization length under the thermodynamic limit in the
localization regime. It avoids the difficulties faced by the
traditional transfer matrix method; and without the sampling
process, the accuracy can be improved easily. As we showed, the
results of our method coincide very well with the known analytical
results of the Lorentzian and the Gaussian distributions, including
the anomalous behaviors at the band center and the band edge. It is
quite efficient when the distribution of diagonal disorder is
nonsingular, especially for moderate disorder. Furthermore, we found
that the Lorentzian distribution should give clues to the anomalies
in the Gaussian distribution. Although it faces some difficulties
for the cases like off-diagonal disorder or correlated disorder,
this method can be generalized to the coupled multichain system with
diagonal uncorrelated disorder.

\begin{acknowledgments}
This work was supported by National Natural Science Foundation of
China, the National Program for Basic Research of MOST of China, and
the Knowledge Innovation Project of Chinese Academy of Sciences.
\end{acknowledgments}

\bibliography{bib}

\end{document}